\documentclass[twocolumn,showpacs,preprintnumbers]{revtex4-1}
\usepackage{amsfonts}
\usepackage{amsmath}
\usepackage{graphicx}
\usepackage{dcolumn}
\usepackage{bm}
\usepackage{epstopdf}
\usepackage{hyperref}
\usepackage{subfigure}
\usepackage[dvipsnames]{xcolor}

\begin{document}

\preprint{}
\title{Traveling wave method for simulating geometric beam coupling impedance of a beamscreen with pumping holes}
\author{Sergey Arsenyev}
\email{sergey.arsenyev@cern.ch}
\author{Alexej Grudiev}
\author{Daniel Schulte}
\affiliation{CERN, Geneva, Switzerland}
\date{\today }

\begin{abstract}
In particle accelerators, pumping holes in a vacuum chamber can be a source of unwanted broadband coupling impedance, leading to beam instabilities.
Analytical methods have been previously developed to estimate the impedance of holes in circular-like chambers e.g. the beamscreen of the Large Hadron Collider (LHC).
More sophisticated chamber designs like that of the High Energy LHC (HE-LHC) and the Future Circular Collider (the hadron-hadron option FCC-hh) call for a different way to calculate the impedance.
We propose using decomposition of the wakefield into synchronous traveling waves and employing a numerical solver to find the impedance of each wave.
This method is compared to the direct time domain wakefield calculation method and its greater sensitivity to small impedances is shown.
\end{abstract}

\maketitle

\section{Introduction}
\label{Intro}

\begin{figure}[!htb]
\centering
\includegraphics[width=90mm, angle=0]{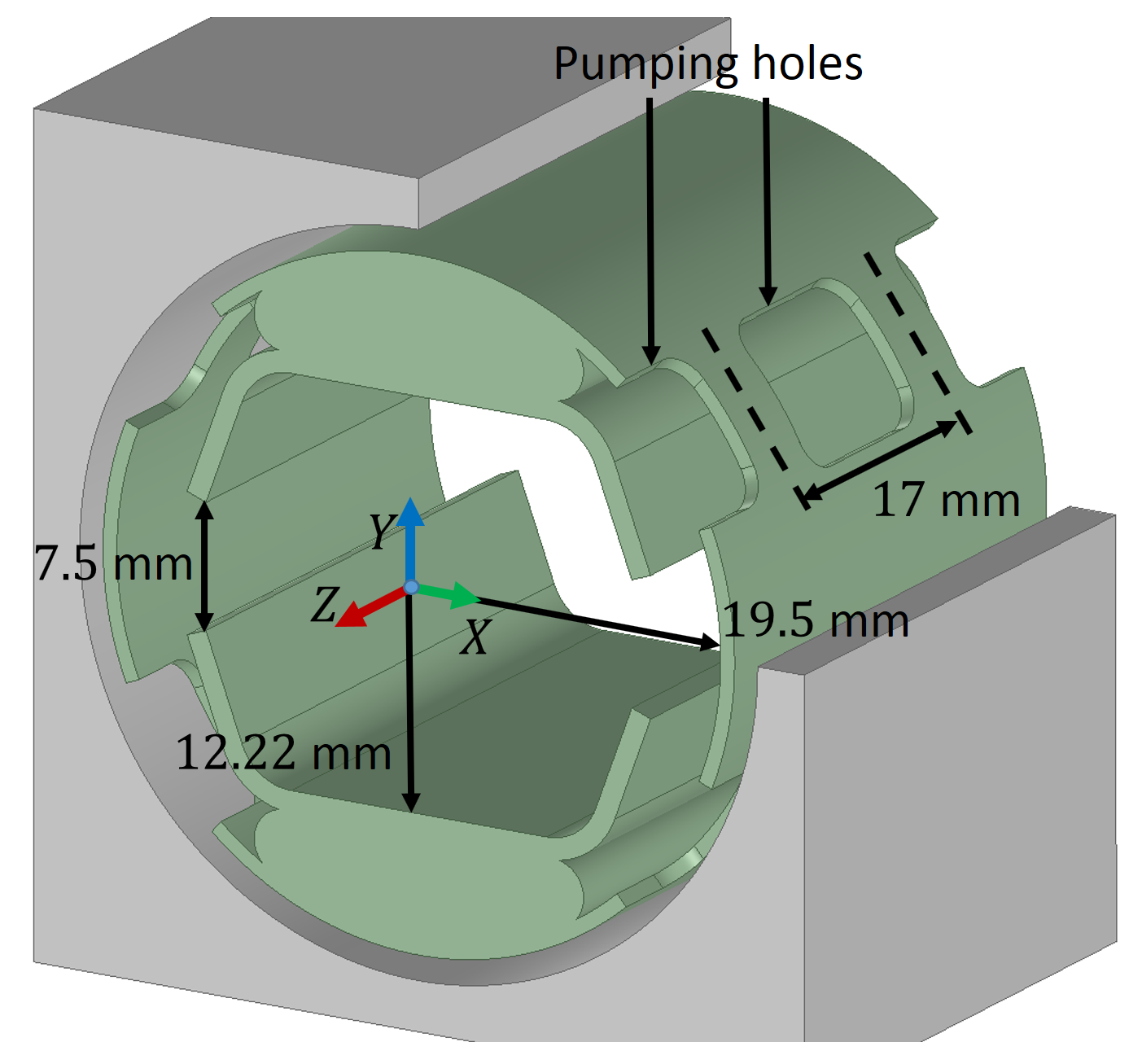}
\caption{FCC-hh beamscreen (green) inside a magnet cold bore (grey). Only the features important to the study are shown, which excludes the cooling channel and the surface coatings. }
\label{FCC_bs}
\end{figure}

In hadron colliders such as the Large Hadron Collider (LHC), the proposed High Energy LHC (HE-LHC) and the Future Circular Collider (the hadron-hadron option FCC-hh), the beamscreen separates the particle beam from the magnet cold bore to avoid having the synchrotron radiation heat load on the cold mass (Fig.~\ref{FCC_bs}).
The pumping holes connect the space inside the beamscreen to the outer region from where the air is pumped out.
In this paper, we propose a new method for calculation of the geometrical beam coupling impedance of the pumping holes.
In particular, we focus on the imaginary part of the longitudinal and the transverse impedances $\mathrm{Im}(Z_{||}(f))$ and $\mathrm{Im}(Z_{\perp}(f))$.
In the mentioned colliders, estimating these quantities at the frequencies of the bunch spectrum (up to $\sim 1$ GHz) is critical to ensure single-bunch beam stability.
Power dissipation through the holes, related to $\mathrm{Re}(Z_{||}(f))$, is another important aspect of the impedance of the holes, but is beyond the scope of this paper.

For a beamscreen with a circular cross-section, analytical models based on Bethe's theory can be applied~\cite{Kurennoy, Gluckstern, DeSantos, Mostacci_holes}.
However, purely analytical models are not applicable if the cross-section of the beamscreen is far from a circle, which is the case for the FCC-hh and the HE-LHC beamscreens due to the additional shielding (Fig.~\ref{FCC_bs}).
A semi-analytical approach suggested in \cite{Kurennoy_95} employs Bethe's formalism together with 2D simulations to estimate the fields at the holes.
However, even this method is not applicable to the FCC-hh and the HE-LHC beamscreens because the holes cannot be considered small and the fields inside a hole vary by two orders of magnitude. 
More than that, even for a circular beamscreen cross-section, theoretical models can be wrong by a factor of a few as will be shown in~\autoref{circ_pipe_sec}.
This error comes from neglecting the interference between consecutive holes and from not perfectly accounting for the influence of the outer chamber.

Therefore, computational tools are needed.
Often, time domain wakefield simulations with codes like CST~\cite{CST}, GdfidL~\cite{GdFidl}, ACE3P~\cite{ACE3P} are used to calculate impedance of complex 3D structures.
However, in the case of a beamscreen with pumping holes, the impedance per unit length can be so low that time domain simulations become not suitable due to the limited numerical accuracy.

The traveling wave method solves this problem by relating the impedance of the holes to a sum over waves traveling in the periodic structure.
This technique by itself is not new and was applied before to estimate the impedance of a periodically corrugated beam pipe~\cite{Mostacci_pipe},~\cite{Bane}.
However, to the best knowledge of the authors, impedance calculation through decomposition into traveling waves was never done for pumping holes (the method was briefly mentioned without a proof in an earlier study by the authors~\cite{IPAC_holes}).
The method can only be applied to beamscreens where the positions of the holes follow a regular pattern.
This is the case for the FCC-hh and the HE-LHC beamscreens, where the positions of the holes are not randomized (like in the LHC) in order to simplify the manufacturing and reduce the cost.
The pattern repeats uninterrupted for the length of a cryo-dipole and part of the interconnect (15.3 m in total) with about 900 periods, making the beamscreen well suited for the approximation of an infinite periodic structure.

An important difference between the implementation of the traveling wave method in~\cite{Mostacci_pipe},~\cite{Bane} (for the case of a corrugated pipe) and the implementation in this paper is that the search for the synchronous waves cannot be purely analytical, but requires eigenmode simulations.
Nevertheless, as a simulation tool, such method offers a valuable advantage over time domain simulations, as it allows to calculate much smaller impedances.
This advantage is related to the fact that the method only requires solving for eigenmodes in a single period of the structure, making the computation much lighter.

\section{Description of the method}
\label{description}

\begin{figure}[!htb]
\centering
\includegraphics[width=90mm, angle=0]{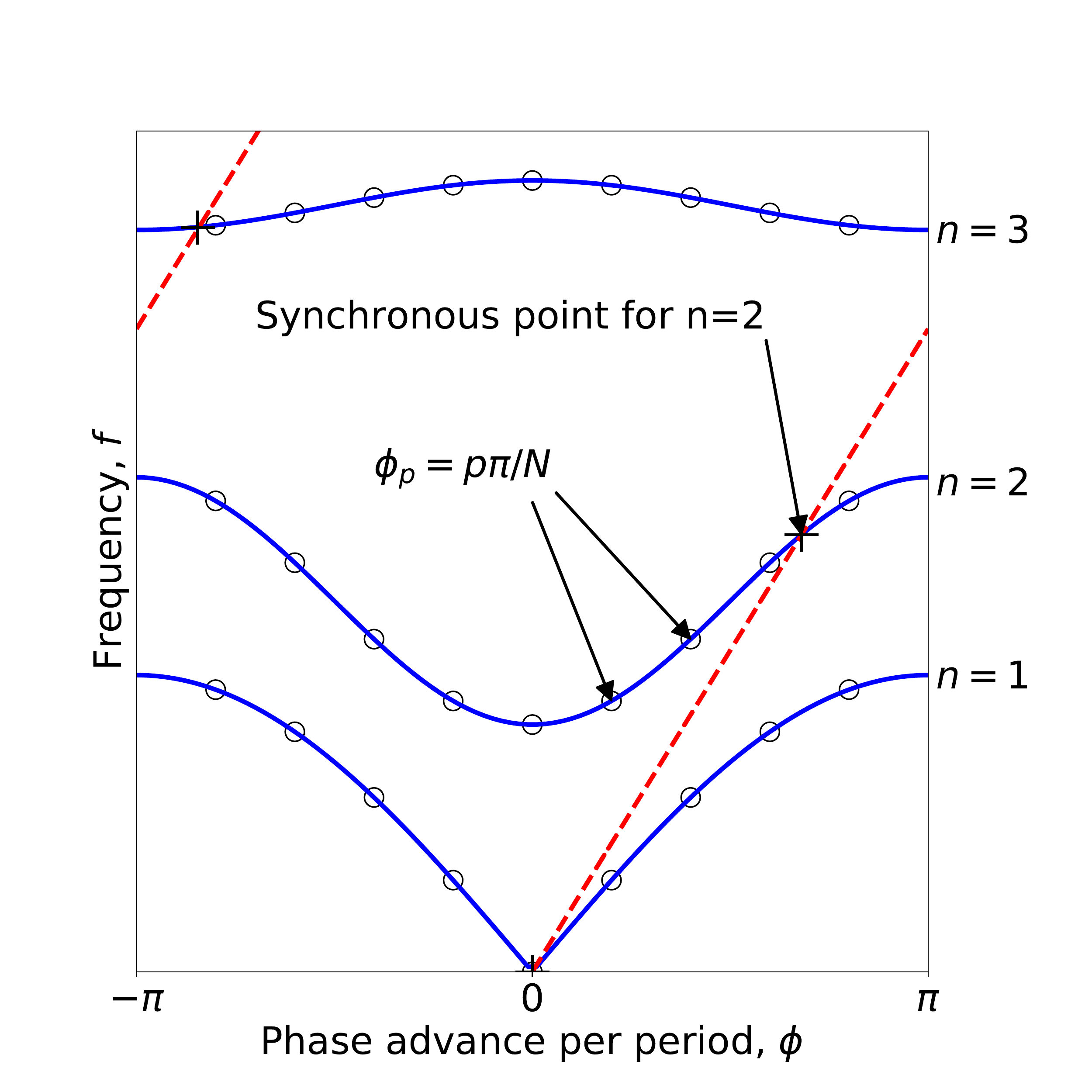}
\caption{An example of the dispersion diagram of a periodic structure with $N$ periods. The blue lines represent the first three bands propagating in the structure, and the red lines represent the synchronous condition with the beam.}
\label{Dispersions_example}
\end{figure}

To employ the method, we first find the traveling waves synchronous with the beam. 
For this, one period of a structure is simulated using an Eigenmode solver, with the phase advance over the period $\phi$ given as a parameter and the eigenmode frequencies $\omega$ read as the output.
The phase advance over one period is scanned over the Brillouin zone to obtain dispersion curves $\omega_n(\phi)$ (blue lines in Fig.~\ref{Dispersions_example}).
We then find intersections ($\phi_n^{syn},~\omega_n^{syn}$) of these curves with the synchronous line of the particle beam.
In this paper the beam is assumed to be ultra-relativistic ($\beta_{rel} \approx 1$) as is the case in FCC-hh and HE-LHC, although this is not a necessary condition to apply the method.
For an ultra-relativistic beam the synchronous line on the dispersion diagram is $\omega=\phi c / L$, appearing as multiple branches in the Brillouin zone (red lines in Fig.~\ref{Dispersions_example}). 
Here $c$ is the speed of light and $L$ is the length of one period of the structure.

Then, for all the synchronous waves ($\phi_n^{syn},~\omega_n^{syn}$), the electromagnetic fields are integrated along the beam path to obtain the longitudinal and the transverse voltage kicks $V_{syn}^{||}$ and $V_{syn}^{\perp}$
\begin{equation}
\begin{split}
{V_{syn}^{||}}_n= \frac{1}{e} \int_0^L F_{||, n}(s, t=s/c) ds, \\
{V_{syn}^{\perp}}_n= \frac{1}{e} \int_0^L F_{\perp, n}(s, t=s/c) ds,
\end{split}
\label{Voltages}
\end{equation}
where $s$ is the longitudinal coordinate, $t$ is the time, and $F_{||, n}$ and $F_{\perp, n}$ are the complex longitudinal and transverse Lorentz forces experienced by a test particle of charge $e$.
These voltages are then used to compute the shunt impedances $(R/Q)$ of the waves.
In this paper we use the ``circuit" definition for shunt impedances:
\begin{equation}
{\left( \frac{R}{Q}\right)_{syn}^{||}}_n=\frac{|{V_{syn}^{||}}_n|^2}{2 \omega_n^{syn} U_n^{syn}},~{\left( \frac{R}{Q}\right)_{syn}^{\perp}}_n=\frac{|{V_{syn}^{\perp}}_n|^2}{2 \omega_n^{syn} U_n^{syn}},
\label{RoverQ}
\end{equation}
where $U_n^{syn}$ is energy stored in one period of the $n$-th synchronous traveling wave. 

Finally, the computed $(R/Q)$ are used to find the longitudinal and the transverse beam coupling impedances per period of the structure $Z_{||}(\omega) / N$ and $Z_{\perp}(\omega) / N$:
\begin{equation}
\begin{split}
\frac{Z_{||}(\omega)}{N} = i \omega \sum_n \alpha_n \frac{1}{\omega_n^{syn}} {\left( \frac{R}{Q}\right)_{syn}^{||}}_n + O(\omega^2), \\
\frac{Z_{\perp}(\omega)}{N} = i \sum_n \alpha_n \frac{\omega_n^{syn}}{c} {\left( \frac{R}{Q}\right)_{syn}^{\perp}}_n + O(\omega), \\
\end{split}
\label{Z_equations}
\end{equation}
where $N$ is the number of periods, factors $\alpha$ are the corrections due to non-zero group velocities $v_g$ of the waves
\begin{equation}
\alpha(v_g) = 
\begin{cases}
(1-v_g/c)^{-1},~\text{if } \omega_{syn} \neq 0 \\
(1-v_g^2/c^2)^{-1},~\text{if } \omega_{syn} = 0, \\
\end{cases}
\label{alpha_equations}
\end{equation}
and the quantity ${\left( R/Q\right)_{syn}^{||}}_n/\omega_n^{syn}$ should be interpreted as $\lim_{\omega' \to 0} \left( R/Q\right)_n^{||} (\omega')/\omega'$ in the case $\omega_n^{syn} = 0$.
The case $\omega_n^{syn} = 0$ appears when the structure allows propagation of TEM-like waves with zero cut-off frequency ($n=1$ band in Fig.~\ref{Dispersions_example}).
It is crucial for a calculation of the longitudinal impedance of the pumping holes, as will be demonstrated in section \ref{circ_pipe_sec}.

The relations~(\ref{Z_equations}) resemble those that can be obtained for ordinary standing wave cavities~\cite{Zotter}, with the exception of the additional factors $\alpha_n$.
The factors $\alpha_n$ are sometimes ignored for similar problems (see, for example,~\cite{Mostacci_pipe}). 
In many cases, this does not result in significant errors because of a small group velocity $v_g \ll c$.
However, in the case of the pumping holes, a consistent disagreement in simulations prompted a strict derivation of the factors (see Appendix).
For the case of non-zero synchronous crossing ($\omega_n^{syn} \neq 0$) the factors were also experimentally observed for CLIC PETS structure \cite{Lars} and later derived in \cite{Walter} and \cite{Bane}.
While taking a different path, the derivation in Appendix leads to the same result for $\omega_n^{syn} \neq 0$ and a new result for the previously not considered case $\omega_n^{syn} = 0$.

\section{Application and benchmarking of the method}

Below we consider three different types of periodic structures: 
\begin{itemize}
\item{simple bellows}
\item{pumping holes in a circular pipe for which analytical models can be applied}
\item{FCC-hh beamscreen}
\end{itemize}

For each structure, the traveling wave analysis is done as described in~\autoref{description}. 
One period of a structure is independently simulated using the Eigenmode solvers of Ansys Electronics Desktop~\cite{HFSS} and the CST Microwave Studio~\cite{CSTMS}, making sure that the results agree.
In both solvers, the tetrahedral meshing method is used to better approximate the curved geometries. 
Since we focus only on the geometrical impedance, all metal walls are considered to be made of a perfect conductor.

A MATLAB~\cite{Matlab} script is used to externally launch the eigenmode solver and to read out the results.
The script scans points along the synchronous line (red line in Fig.~\ref{Dispersions_example}) and finds all the intersections with the mode dispersion curves ($\phi_n^{syn},~f_n^{syn}$) in the desired frequency range. 
Finding the intersection points consumes most of the computational time. 
Once the points are found, the longitudinal and the transverse shunt impedances are calculated by integrating the electromagnetic force along the beam path.
Since all the considered structures possess mirror symmetries in $X$ and $Y$, symmetry planes are applied: the magnetic symmetries in both $XZ$ and $YZ$ for the longitudinal impedance, or the magnetic symmetry in $XZ$ and the electric symmetry in $YZ$ for the transverse impedance in the $X$-direction. 
In all considered structures the transverse impedance in the $Y$-direction is lower than (or equal to) the transverse impedance in the $X$-direction and is not discussed.

The traveling wave method is compared to the Wakefield solver of CST~\cite{CST} - a well-established time domain solver for wake and impedance calculations. 
The impedance computed in the CST Wakefield solver is a result of a direct Fourier transformation of the wake function, with both the longitudinal and the transverse impedances measured in $\Omega$.
We normalize the transverse impedance by the transverse offset to convert the result to $\Omega / m$ which allows for the comparison.
For the wakefield computation, a 10-period long structure is terminated with open boundary conditions at the ends. 
Since the open boundary does not represent an infinitely repeating structure, reflected waves from the ends produce an unwanted contribution to the impedance. 
To cancel out this contribution, a 20-period long structure is simulated and the difference between 10 and 20 periods was considered for impedance calculation. 
Care is taken to make sure that this difference is reproduced when 30 periods are compared to 20 periods.

\subsection{Bellows}
\begin{figure}[!htb]
\centering
\includegraphics[width=70mm, angle=0]{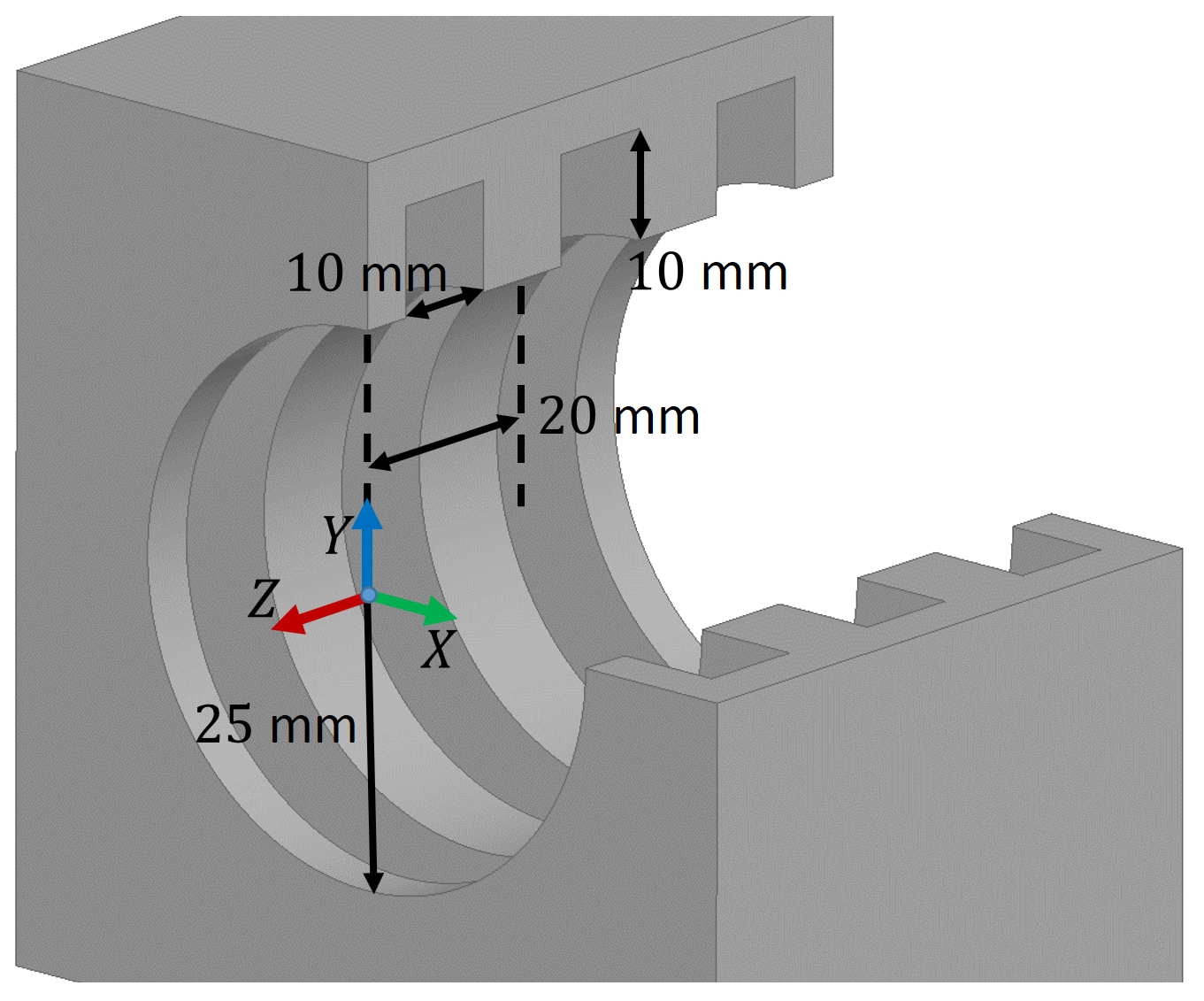}
\caption{The simulated bellows structure (3 periods are shown).}
\label{Bellows_sketch}
\end{figure}

\begin{figure}[!htb]
\centering
\includegraphics[width=90mm, angle=0]{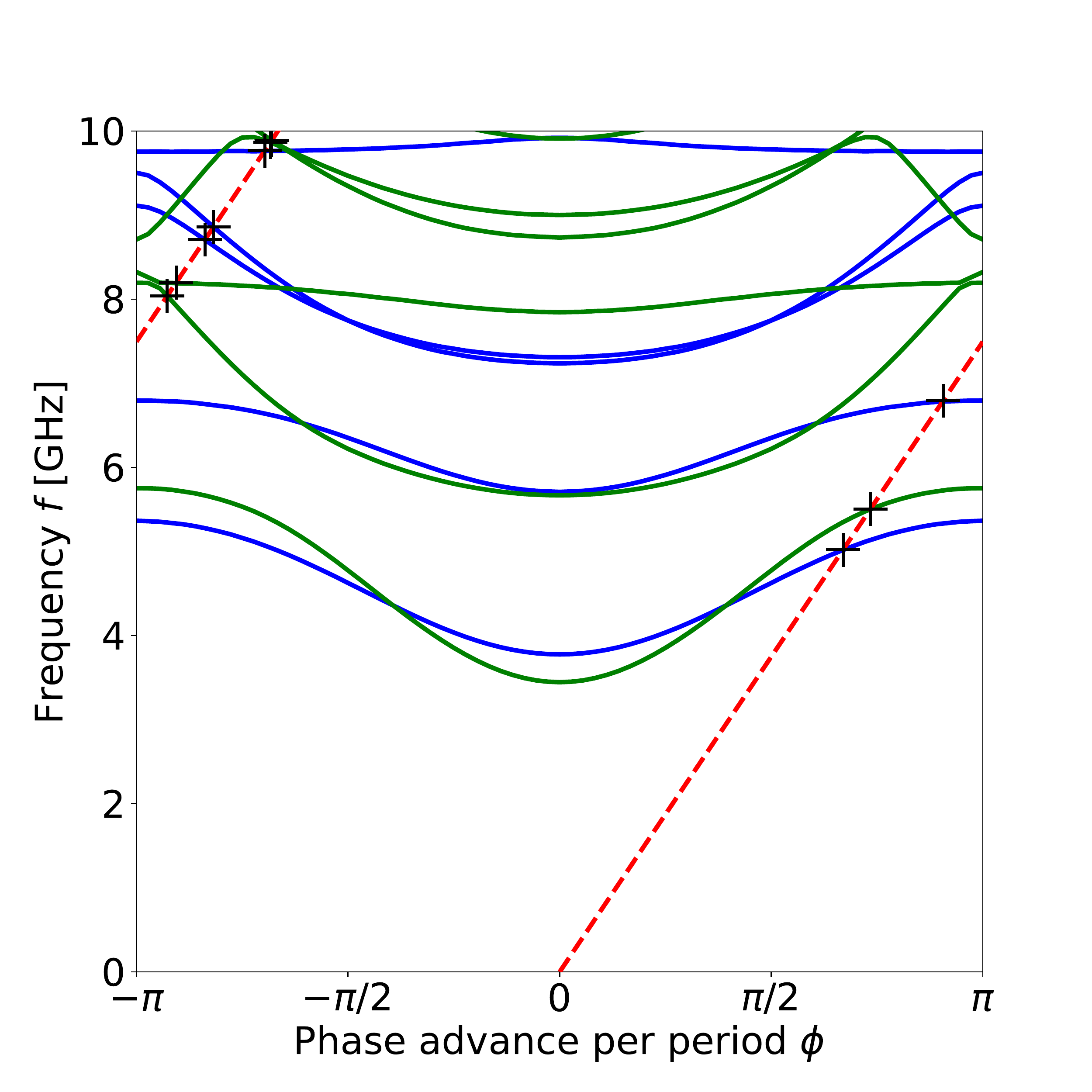}
\caption{Dispersion diagram for the bellows structure depicted in Fig.~\ref{Bellows_sketch}. The blue lines correspond to longitudinal modes, and the green lines correspond to transverse modes. The synchronous points are marked with black crosses.}
\label{Dispersions_bellows}
\end{figure}

The bellows structure shown in Fig.~\ref{Bellows_sketch} is one of the simplest geometries to simulate, hence a good agreement between the different methods is expected.
Dispersion curves for the first five longitudinal and transverse modes are shown in Fig.~\ref{Dispersions_bellows}.
Results of the traveling wave method are shown in Tables~\ref{Bellows_long} and \ref{Bellows_trans} for the longitudinal and the transverse cases, respectively.
The corresponding total impedances given by Eq.~\ref{Z_equations} agree within $5\%$ with the results of the CST wakefield solver (also shown in Tables~\ref{Bellows_long} and \ref{Bellows_trans}).
For this simple structure, the impedances are dominated by the lowest frequency traveling wave in both the longitudinal and the transverse cases. 
Notice also that if the correction factors $\alpha$ were not taken into account, the results would have been underestimated by 40\%.

\begin{table}[hbt]
\centering
\caption{Longitudinal traveling wave data for one period of the bellows structure depicted in Fig.~\ref{Bellows_sketch}. The final result for $\mathrm{Im} \left( Z_{||} \right) / f$ is the sum of the elements of the third column weighted with the corresponding factors $\alpha$. The result of the wakefield solver is also shown for comparison.}
\begin{ruledtabular}
\begin{tabular}{lccr} 
Mode number & $f_{syn}$ [GHz] & $\frac{1}{f_{syn}} \big[ \frac{R}{Q}\big]_{syn}^{||}~[\frac{\Omega}{\mathrm{GHz}}]$ & $\alpha$ \\
\hline
1 & 5.03 & $2.37\times10^0$ & 1.37\\
2 & 6.79 & $<10^{-5}$ & 1.04\\
3 & 8.71 & $9.15\times10^{-2}$ & 0.68\\
4 & 8.86 & $<10^{-5}$ & 0.63\\
5 & 9.77 & $<10^{-5}$ & 1.00\\
\hline
\multicolumn{2}{l}{$\mathrm{Im} \left( Z_{||} \right) / f$ (traveling wave)}  & $3.31~\Omega/ \mathrm{GHz}$ & \\
\multicolumn{2}{l}{$\mathrm{Im} \left( Z_{||} \right) / f$ (wakefield solver)} & $3.50~\Omega/ \mathrm{GHz}$ & \\
\end{tabular}
\end{ruledtabular}
\label{Bellows_long}
\end{table}

\begin{table}[hbt]
\centering
\caption{Transverse traveling wave data for one period of the bellows structure depicted in Fig.~\ref{Bellows_sketch}. The final result for $\mathrm{Im} \left( Z_x \right)$ is the sum of the elements of the third column weighted with the corresponding factors $\alpha$. The result of the wakefield solver is also shown for comparison.}
\begin{ruledtabular}
\begin{tabular}{lccr} 
Mode number & $f_{syn}$ [GHz] & $\frac{2 \pi f_{syn}}{c} \big[ \frac{R}{Q} \big]_{syn}^x ~[\frac{\Omega}{m}]$ & $\alpha$ \\
\hline
1 & 5.51 & $2.85\times10^2$ & 1.38\\
2 & 8.04 & $5.18\times10^0$ & 0.58\\
3 & 8.20 & $1.31\times10^{-5}$ & 0.99\\
4 & 9.87 & $9.12\times10^0$ & 0.79\\
5 & 9.89 & $1.82\times10^{-3}$ & 0.69\\
\hline
\multicolumn{2}{l}{$\mathrm{Im} \left( Z_x \right)$ (traveling wave)} & $403~\Omega/m$ & \\
\multicolumn{2}{l}{$\mathrm{Im} \left( Z_x \right)$ (wakefield solver)} & $410~\Omega/m$ & \\
\end{tabular}
\end{ruledtabular}
\label{Bellows_trans}
\end{table}

\subsection{Pumping holes in a circular pipe}
\label{circ_pipe_sec}

\begin{figure}[!htb]
\centering
\includegraphics[width=70mm, angle=0]{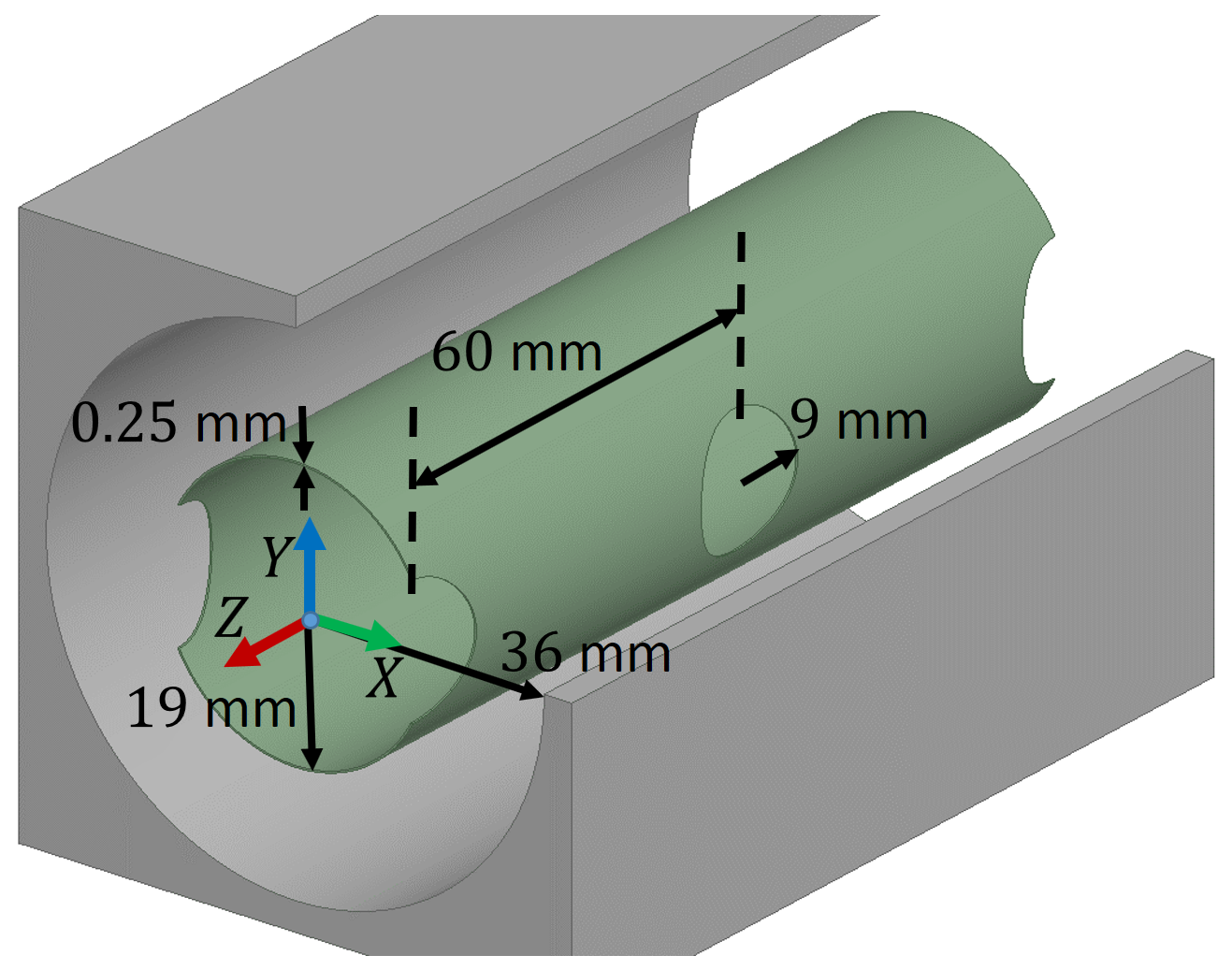}
\caption{Pumping holes in a circular pipe for the case $M=2$ (2 periods are shown).}
\label{Circular_pipe_sketch}
\end{figure}

We now consider a simple structure with pumping holes, shown in Fig.~\ref{Circular_pipe_sketch}.
It consists of a circular pipe with two circular holes per cross section, opening up to a cylindrical outer region. 
This simple geometry allows comparing calculated impedances to an analytical model by S. Kurennoy~\cite{Kurennoy}. 
According to this model (and confirmed by the later study accounting for the outer coaxial region~\cite{DeSantos}), for circular holes the imaginary part of impedances are given by
\begin{gather}
-\frac{\mathrm{Im}(Z_{||})}{N} = \frac{Z_0 R_h^3 M}{6 \pi^2 c b^2} \omega\\
-\frac{\mathrm{Im}(Z_{\perp})}{N} = 
\begin{cases}
\frac{2 Z_0 R_h^3 M}{3 \pi^2 b^4},~\text{if } M=1~\text{or}~2 \\
\frac{Z_0 R_h^3 M}{3 \pi^2 b^4},~\text{if } M \geq 3,
\end{cases}
\label{circpipe_theory}
\end{gather}
where $N$ is the number of periods, $Z_0 = 377~\Omega$ is the impedance of vacuum, $R_h$ is the hole radius, $b$ is the inner radius of the pipe, and $M$ is the number of holes per cross section ($M=2$ in the considered structure).
For the considered geometry, the effect of the wall thickness on the polarizabilities can be neglected (see Fig.~5 of ~\cite{McDonald}).

\begin{figure}[!htb]
\centering
\includegraphics[width=90mm, angle=0]{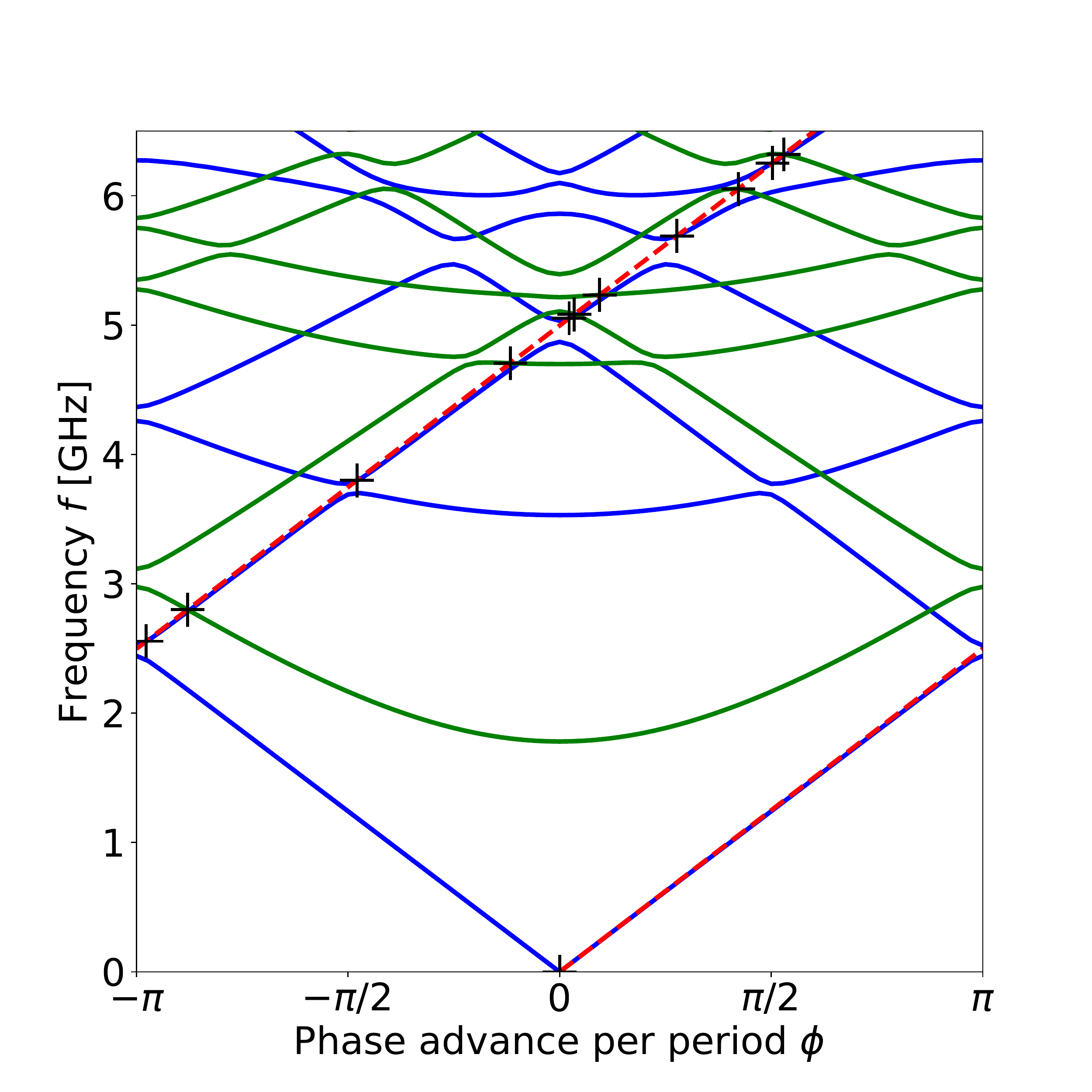}
\caption{Dispersion diagram for the circular pipe structure depicted in Fig.~\ref{Circular_pipe_sketch}. The blue lines correspond to longitudinal modes, and the green lines correspond to transverse modes ($X$-direction). The synchronous points are marked with black crosses.}
\label{Dispersions_circ_pipe}
\end{figure}

An important complication relative to the case of bellows arises from the fact that the outer region allows propagation of a TEM-like mode, similar to a coaxial line.
This mode is represented by the lowest blue line in the corresponding dispersion diagram (Fig.~\ref{Dispersions_circ_pipe}), almost matching the synchronous line.
The exact intersection with the synchronous line appears at the point $\phi=0,~f=0$ and is treated as a zero synchronous crossing as described in section~\ref{description}.

Results of the traveling wave method are shown in Tables~\ref{Circ_pipe_long} and \ref{Circ_pipe_trans} for the longitudinal and the transverse cases, respectively.
In case of the longitudinal impedance, the contribution of the first synchronous wave is much greater than that of all the other modes combined.
This is due to the very high factor $\alpha$ caused by the dispersion curve of the first band almost matching the synchronous line.
This shows the importance of treating the case $\omega_{syn} = 0$ separately.
Indeed, if we had mistakenly used the non-zero crossing expression for $\alpha$ (Eq.~(\ref{alpha_equations}), top), the final answer would have been a factor of 2 off.

\begin{figure}[!htb]
\centering
\includegraphics[width=90mm, angle=0]{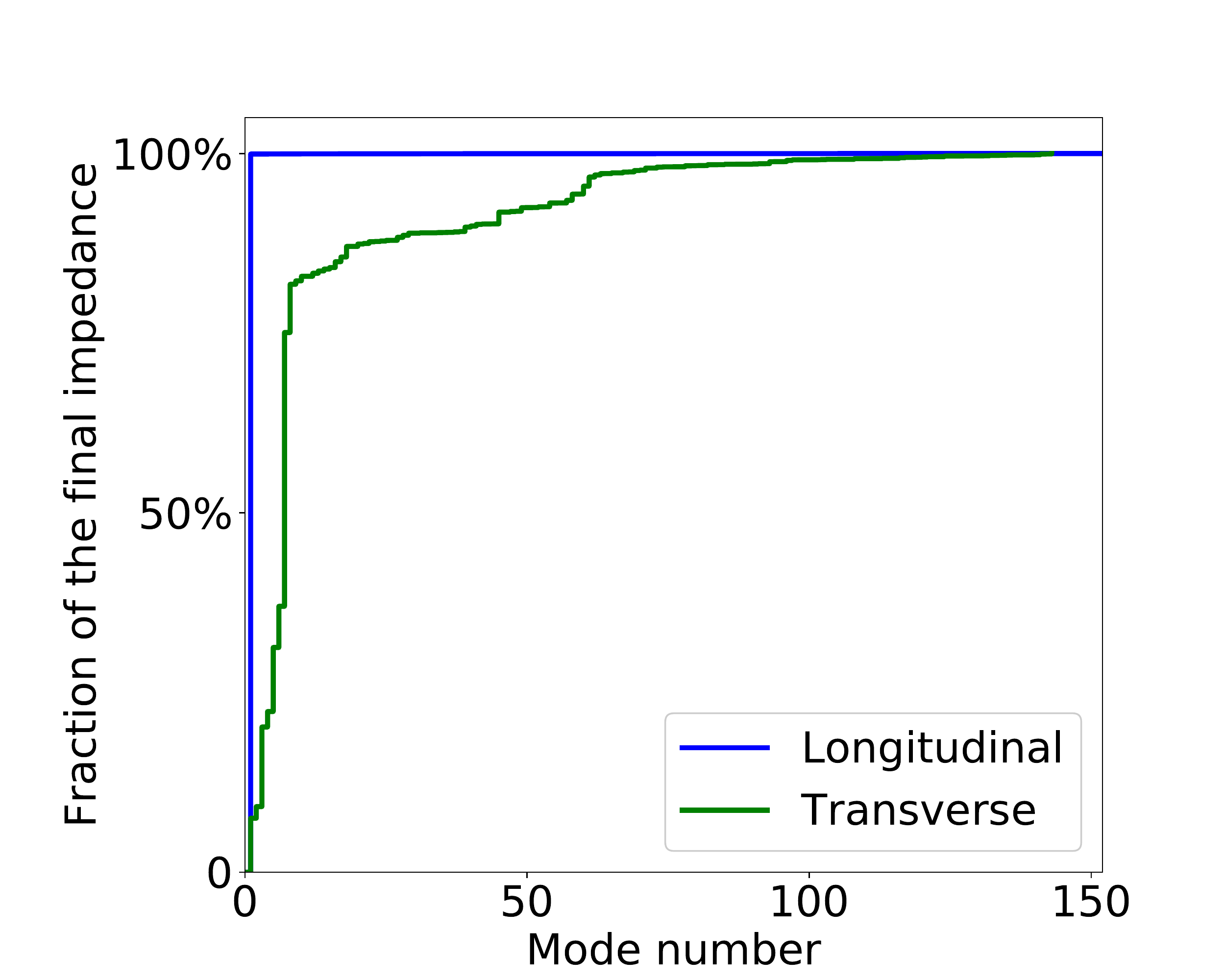}
\caption{Convergence check for the circular pipe structure depicted in Fig.~\ref{Circular_pipe_sketch}.}
\label{Convergence_circ_pipe}
\end{figure}

In case of the transverse impedance, the contributions are spread over a large number of traveling waves.
Modes up to the frequency of 20 GHz were taken into account, with 143 modes in total.
To check that the number of modes is sufficient, the evolution of the impedance is plotted as a function of the number of modes in Fig.~\ref{Convergence_circ_pipe}.
After summing more than 50 traveling waves ($f > 15$ GHz), the impedance converges to a value which is taken as the total impedance.

\begin{table}[hbt]
\centering
\caption{Longitudinal traveling wave data for one period of the circular pipe structure depicted in Fig.~\ref{Circular_pipe_sketch}. The final result for $\mathrm{Im} \left( Z_{||} \right) / f$ is the sum of the elements of the third column weighted with the corresponding factors $\alpha$. The results of the wakefield solver and the analytical formula are also shown for comparison.}
\begin{ruledtabular}
\begin{tabular}{lccr} 
Mode number & $f_{syn}$ [GHz] & $\frac{1}{f_{syn}} \big[ \frac{R}{Q}\big]_{syn}^{||}~[\frac{\Omega}{\mathrm{GHz}}]$ & $\alpha$ \\
\hline
1 & 0.00 & $5.77\times10^{-3}$ & 88.1\\
2 & 2.54 & $<10^{-5}$ & 3.50\\
3 & 3.79 & $<10^{-5}$ & 2.44\\
4 & 5.05 & $<10^{-5}$ & 1.94\\
5 & 5.69 & $1.31\times10^{-5}$ & 2.09\\
... & & & \\
150 & 19.96 & $<10^{-5}$ & 0.88 \\
151 & 19.97 & $<10^{-5}$ & 0.61 \\
152 & 19.98 & $<10^{-5}$ & 0.72 \\
\hline
\multicolumn{2}{l}{$\mathrm{Im} \left( Z_{||} \right) / f$ (traveling wave)} & $0.508~\Omega/ \mathrm{GHz}$ & \\
\multicolumn{2}{l}{$\mathrm{Im} \left( Z_{||} \right) / f$ (wakefield solver)} & $0.52~\Omega/ \mathrm{GHz}$ & \\
\multicolumn{2}{l}{$\mathrm{Im} \left( Z_{||} \right) / f$ (analytical)} & $0.538~\Omega/ \mathrm{GHz}$ &
\end{tabular}
\end{ruledtabular}
\label{Circ_pipe_long}
\end{table}

\begin{table}[hbt]
\centering
\caption{Transverse traveling wave data for one period of the circular pipe structure depicted in Fig.~\ref{Circular_pipe_sketch}. The final result for $\mathrm{Im} \left( Z_x \right)$ is the sum of the elements of the third column weighted with the corresponding factors $\alpha$. The results of the wakefield solver and the analytical formula are also shown for comparison.}
\begin{ruledtabular}
\begin{tabular}{lccr} 
Mode number & $f_{syn}$ [GHz] & $\frac{2 \pi f_{syn}}{c} \big[ \frac{R}{Q} \big]_{syn}^x ~[\frac{\Omega}{m}]$ & $\alpha$ \\
\hline
1 & 2.80 & $3.21\times10^{1}$ & 0.58\\
2 & 4.71 & $4.18\times10^{0}$ & 0.96\\
3 & 5.08 & $3.91\times10^{1}$ & 0.70\\
4 & 5.23 & $4.90\times10^{0}$ & 1.08\\
5 & 6.05 & $2.57\times10^{1}$ & 0.86\\
... & & & \\
141 & 19.89 & $1.54\times10^{-1}$ & 1.44\\
142 & 19.92 & $2.65\times10^{-2}$ & 0.73\\
143 & 19.96 & $1.34\times10^{-1}$ & 1.13 \\
\hline
\multicolumn{2}{l}{$\mathrm{Im} \left( Z_x \right)$ (traveling wave)} & $247~\Omega/m$ & \\
\multicolumn{2}{l}{$\mathrm{Im} \left( Z_x \right)$ (wakefield solver)} & $275~\Omega/m$ & \\
\multicolumn{2}{l}{$\mathrm{Im} \left( Z_x \right)$ (analytical)} & $285~\Omega/m$ & \\
\end{tabular}
\end{ruledtabular}
\label{Circ_pipe_trans}
\end{table}

In Tables~\ref{Circ_pipe_long} and \ref{Circ_pipe_trans}, the results of the traveling wave calculation are also compared to those obtained by the wakefield solver and by the analytical model of Eq.~\ref{circpipe_theory}.
All three methods agree within 15\%.
However, while the agreement between the traveling wave method and the wakefield solver is always to be expected, the agreement with the analytical model is only achieved thanks to the careful choice of the geometry parameters.
In particular, a large distance between the holes $L$ and a large radius of the outer wall $R_{out}$ are necessary for Eq.~\ref{circpipe_theory} to correctly estimate the impedance.

To prove this statement, we consider an example of a geometry that better reflects the features of a real beamscreen, and call it ``Geometry B" (as opposed to the already analyzed ``Geometry A").
In particular, we reduce the period length $L$ to 20 mm to make the holes more tightly packed, and reduce the outer chamber radius $R_{out}$ to 22 mm to reflect the fact that the beamscreen occupies most of the cold bore space.
As far as the analytical approach is concerned, the changes in $L$ and $R_{out}$ have no impact on the impedance.
This, however, is not the case, as is shown in  Table~\ref{two_geom}, where the results of the two computational methods are compared to the analytical model for both geometries.
One can see that the analytical approach works well only for Geometry A, but significantly overestimates the impedance for Geometry B. 
Thus, even if the FCC-hh beamscreen had a circular-like cross-section, the theory could have been only used to get a rough (within a factor of a few) estimate.

\begin{table}[hbt]
\centering
\caption{Comparison between the analytical model and the computational methods for two different geometries. 
The values for $\mathrm{Im} \left( Z_{||} \right)/f$ are in $\Omega/\mathrm{GHz}$ and the values for $\mathrm{Im} \left( Z_x \right)$ are in $\Omega / m$.}
\begin{ruledtabular}
\begin{tabular}{lcccc} 
& \multicolumn{2}{c}{Geometry A} & \multicolumn{2}{c}{Geometry B}\\
Method & $\mathrm{Im} \left( Z_{||} \right)/f$ & $\mathrm{Im} \left( Z_x \right)$ & $\mathrm{Im} \left( Z_{||} \right) / f$ & $\mathrm{Im} \left( Z_x \right)$ \\
\hline
Trav. waves & 0.508 & 247 & 0.163 & 77 \\
Wake. solver & 0.52 & 275 & 0.165 & 87 \\
Analytical & 0.538 & 285 & 0.538 & 285 \\
\end{tabular}
\end{ruledtabular}
\label{two_geom}
\end{table}

\subsection{FCC-hh beamscreen}
\label{FCCbs_sec}
We consider the 2018 version of the FCC-hh beamscreen design (Fig.~\ref{FCC_bs}), which is different from the previous designs considered, for example, in~\cite{IPAC_holes}.
An important difference from the case of the circular pipe (\autoref{circ_pipe_sec}) comes from the addition of the hole shielding.
Because of the shielding, the holes are not directly visible to the beam and instead are coupled to the beam region through a slit.
The narrower the slit, the stronger the shielding of the holes, and, consecutively, the lower the hole impedance.
The low impedance \emph{per period} of the structure, however, still needs an exact estimate, as with 4.8 million periods in the FCC-hh it can reach a critical level.
When estimating low impedances, the traveling wave method is expected to have a significant advantage over the wakefield solver.
This is due to the fact that only one period of the structure needs to be simulated, and due to the more accurate conformal tetrahedral mesh available in eigenmode solvers.

\begin{figure}[!htb]
\centering
\includegraphics[width=90mm, angle=0]{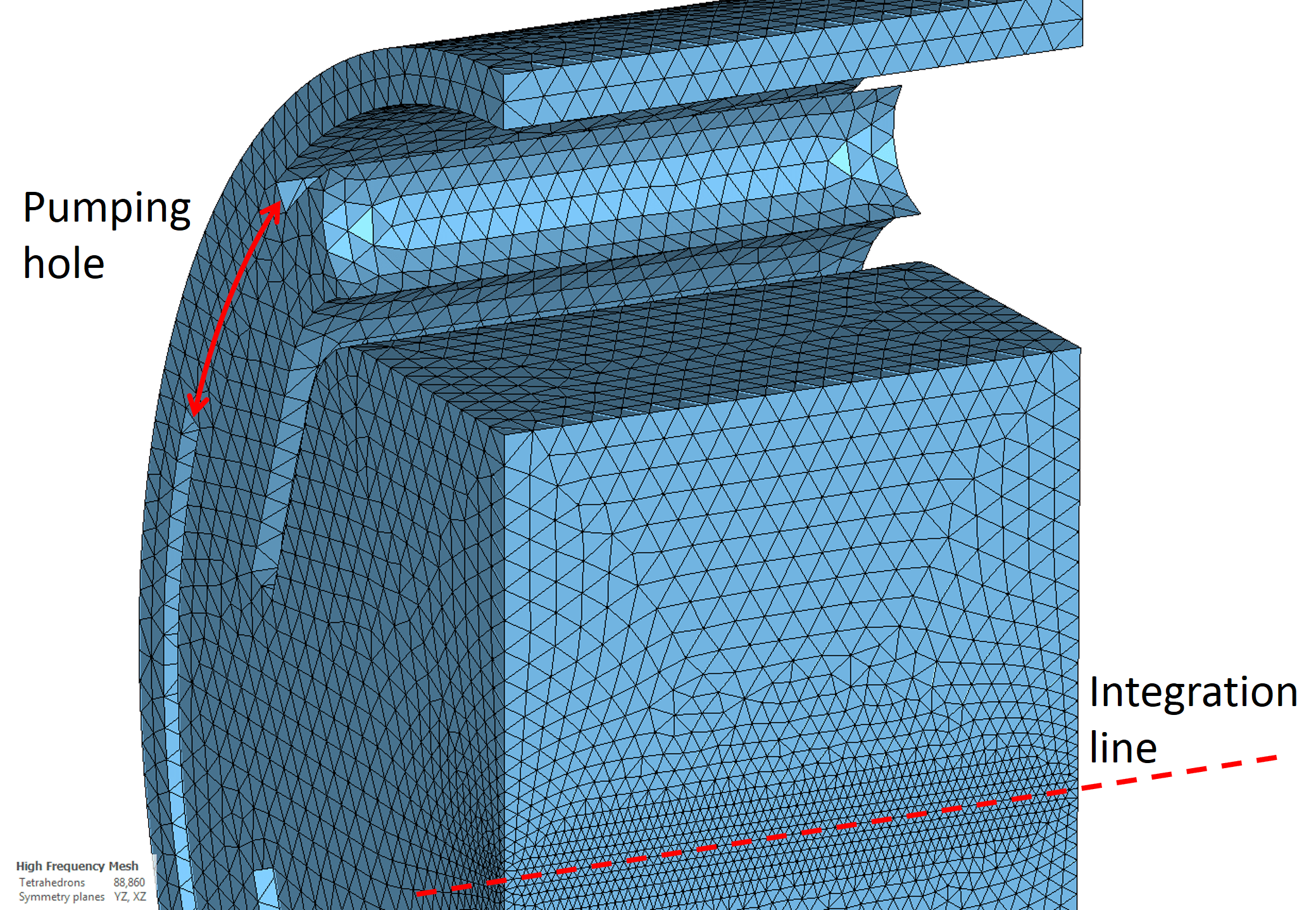}
\caption{Meshing of the vacuum volume in the FCC-hh beamscreen used in the CST eigenmode simulations. 
To calculate very low impedances of the structure with the slit width of 7.5 mm, we used 88860 non-curved tetrahedral meshcells, a 1.1 mesh equilibrate ratio, the 3rd order solver, and 2 adaptive mesh refinement passes. }
\label{mesh}
\end{figure}

In order to calculate very low impedances with the traveling wave method, special care is taken to reduce the numerical noise floor of the eigenmode simulations.
For that, the mesh nodes are forced to lie on the voltage integration line, and the mesh steps along the line are made much smaller than in the rest of the volume (Figure~\ref{mesh}).
The synchronicity condition is strictly enforced, as the phase advance is adjusted until the frequencies of the traveling waves to lie within 10 kHz from the synchronous line.
To measure the noise level, we simulate a modified geometry for which the impedance is known to be strictly zero.
For that, the consecutive holes are connected together creating an infinitely long continous hole.
The cross-section of such structure is a constant of the longitudinal coordinate, leading to the zero geometrical impedance.
Therefore, for each traveling wave, a small numerically obtained impedance gives the noise floor.
%With the mentioned techniques, the noise floor per traveling wave was brought down to $\mathrm{Im} (Z_{||}) / f \leq 10^{-14}~\Omega / \mathrm{GHz}$ and $\mathrm{Im} (Z_{\perp}) \leq 10^{-7}~\Omega / m$.
The mesh and eigenmode solver parameters were chosen to minimize this noise floor.

\begin{figure}[!htb]
\centering
\includegraphics[width=90mm, angle=0]{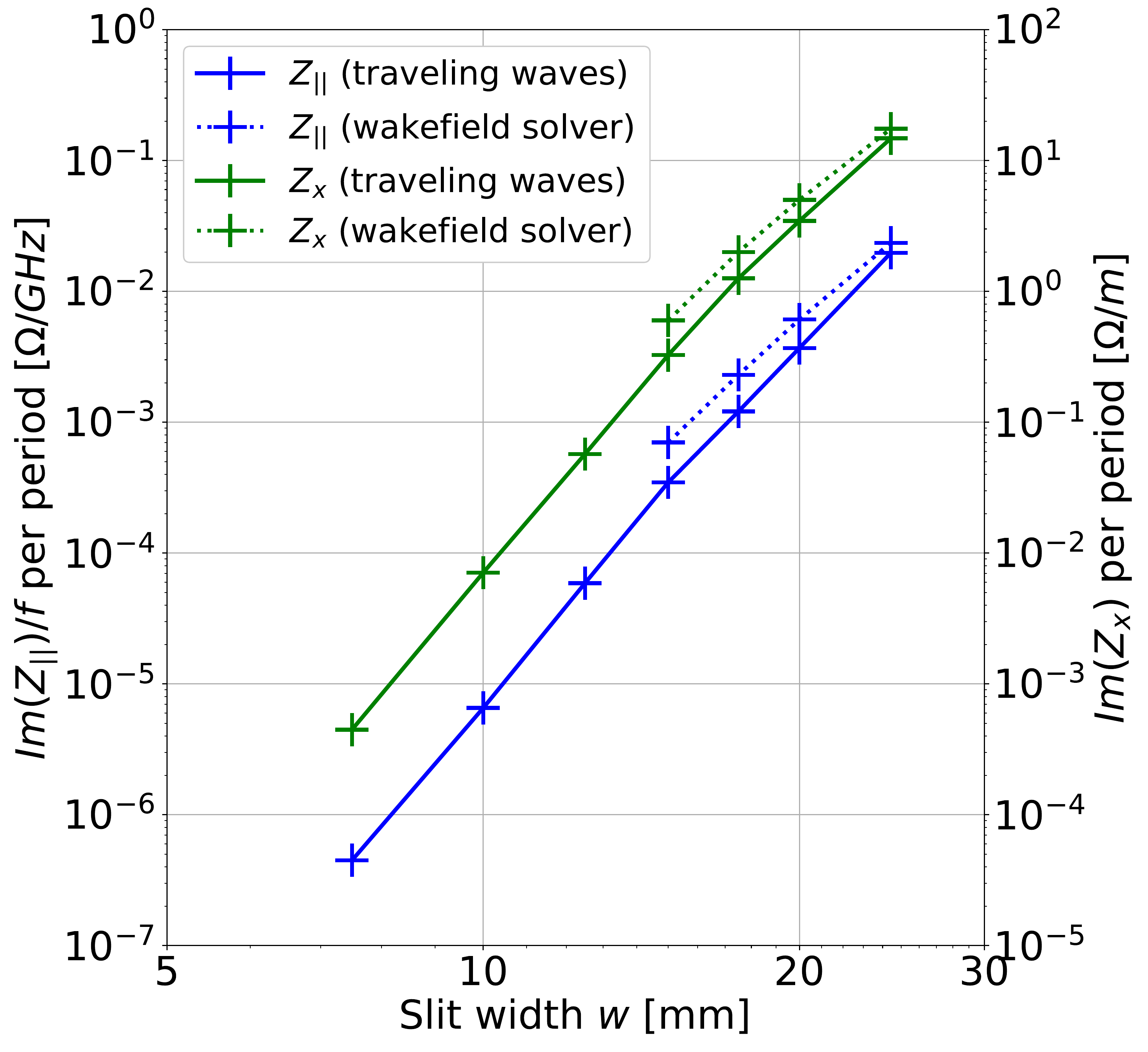}
\caption{Impedance of the holes in the FCC-hh beamscreen as a function of the slit width.}
\label{Imp_vs_slit}
\end{figure}

For the actual FCC-hh beamscreen design, the wakefield solver fails to give an impedance estimate as the impedance is far below the computational noise.
Nevertheless, the wakefield solver can be compared to the traveling wave method for geometries with wider slits, where the impedance is not low.
Such comparison is shown in Fig.~\ref{Imp_vs_slit} for the slit width $w$ ranging from 7.5 mm (the actual design) to 24.44 mm (the screen is completely removed).
For the case of $w = 24.44~\mathrm{mm}$ the two methods agree within 20\%, but the corresponding lines start diverging for narrower slits due to the wakefield results affected by the computational noise.
In the range $w < 15~\mathrm{mm}$ the wakefield solver becomes unusable, while the traveling wave solver gives converging results even for $w = 7.5~\mathrm{mm}$.
Thus, the sensitivity of the traveling wave method to low impedances exceeds the one of the wakefield solver by three orders of magnitude.

\begin{figure}[!htb]
\centering
\includegraphics[width=90mm, angle=0]{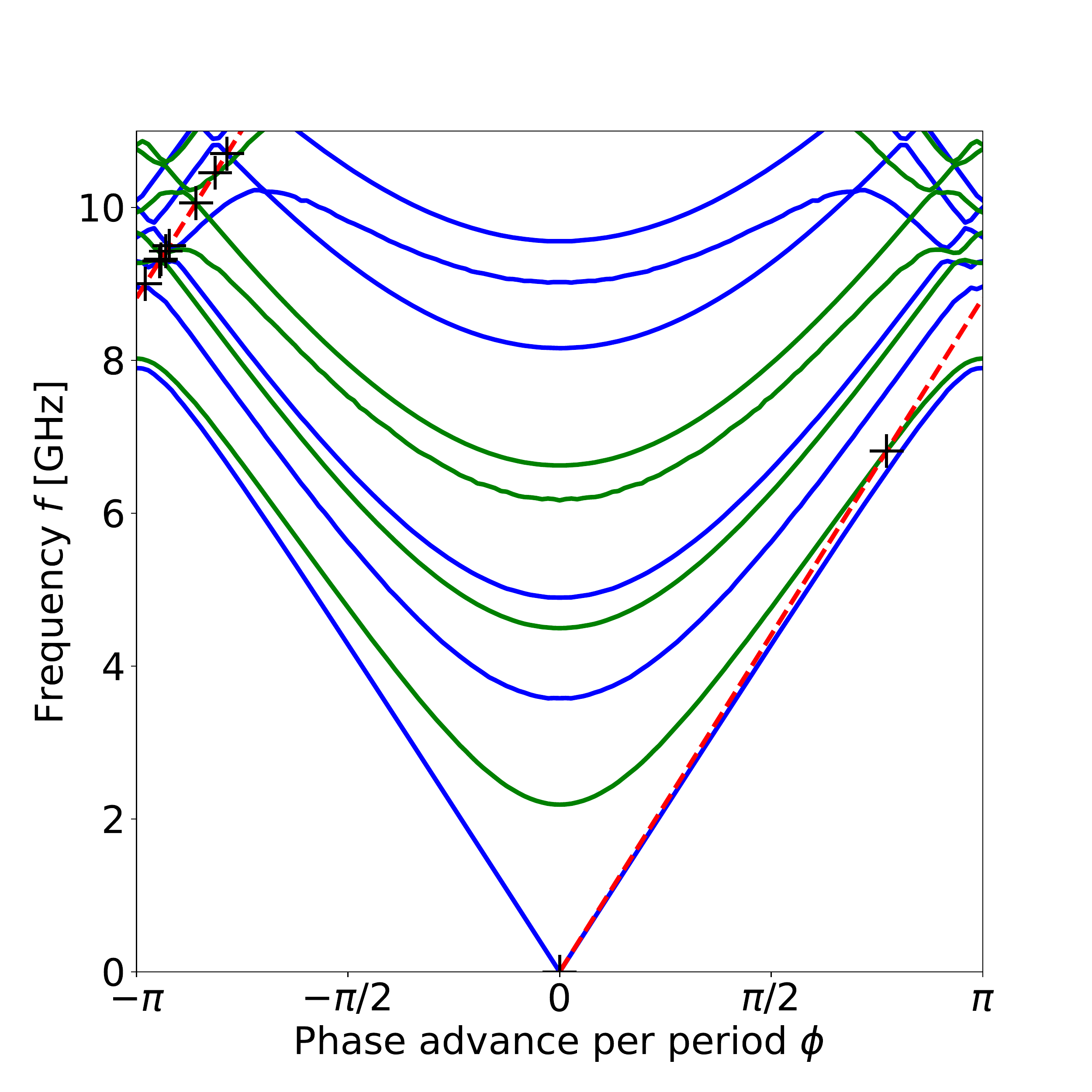}
\caption{Dispersion diagram for the FCC-hh beamscreen with the slit width of 7.5 mm. The green lines correspond to longitudinal modes, and the blue lines correspond to transverse modes ($X$-direction).  The synchronous points are marked with black crosses.}
\label{Dispersions_FCC_7p5mm}
\end{figure}

\begin{figure}[!htb]
\centering
\includegraphics[width=90mm, angle=0]{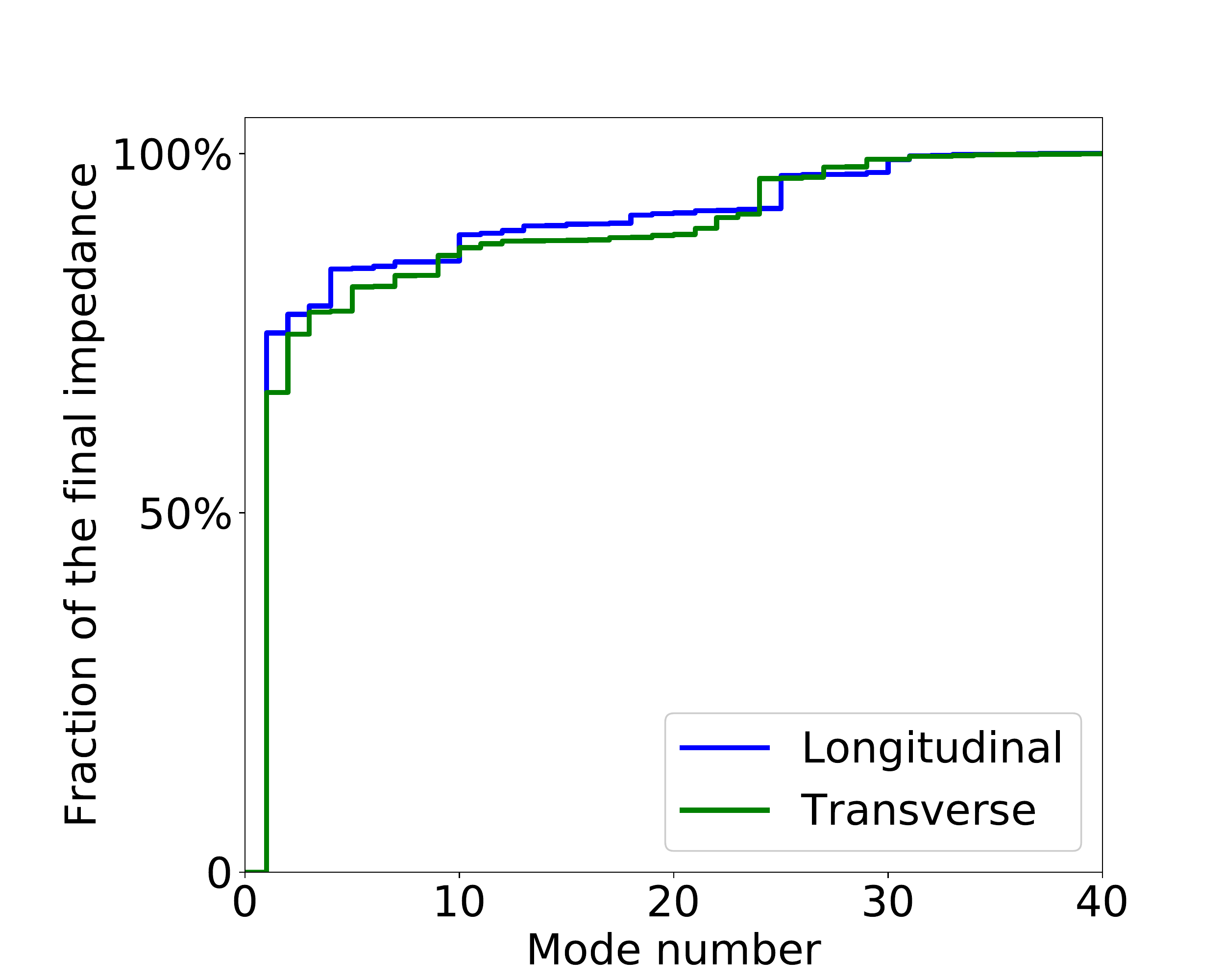}
\caption{Convergence check for the FCC-hh beamscreen with the slit width 7.5 mm.}
\label{Convergence_fcc_7p5mm}
\end{figure}

For narrow slits $w < 15~\mathrm{mm}$, the curves for $\mathrm{Im}(Z_{||})(w)$ and $\mathrm{Im}(Z_x)(w)$ approach straight lines on the log-log scale (Fig.~\ref{Imp_vs_slit}). 
These lines correspond to a very steep dependence on the slit width that can be approximated by $\mathrm{Im}(Z) \propto w^{9.5}$, confirming that the shielding is very effective at reducing the impedance of the holes.
For the actual design ($w = 7.5~\mathrm{mm}$) the traveling wave method gives
\begin{equation}
\begin{split}
\mathrm{Im}(Z_{||})_{per~period} / f = 4.49 \times 10^{-7}~\Omega / \mathrm{GHz} \\
\mathrm{Im}(Z_x)_{per~period} = 4.48 \times 10^{-4}~\Omega / m.
\end{split}
\end{equation}
The corresponding dispersion diagram and the convergence check are plotted in Figures~\ref{Dispersions_FCC_7p5mm} and~\ref{Convergence_fcc_7p5mm}, and the eigenmode data is presented in Tables~\ref{FCC_7p5mm_long} and ~\ref{FCC_7p5mm_trans} for the first 40 modes. 
The total impedance of 4.8 million periods of the holes is
\begin{equation}
\begin{split}
\mathrm{Im}(Z_{||})_{total} / n = 6.7 \times 10^{-6}~\Omega \\
\mathrm{Im}(Z_x)_{total} = 2.2 \times 10^{3}~\Omega / m,
\end{split}
\end{equation}
where, as often used for stability studies, the longitudinal impedance is normalized over $n = f / f_{rev}$, and the revolution frequency $f_{rev} = 3.067~ \mathrm{kHz}$.
In the FCC-hh, the instability threshold for $\mathrm{Im}(Z_{||}) / n$ is in the $\mathrm{m} \Omega$ range~\cite{FCC_short_CDR} and the instability threshold for $\mathrm{Im}(Z_x)$ is in the $\mathrm{M} \Omega /m$ range~\cite{Impedance_budget}.
Therefore, as far as the beam stability is concerned, the impedance of the holes in the FCC-hh beamscreen is completely negligible.

\begin{table}[hbt]
\centering
\caption{Longitudinal traveling wave data for one period of the FCC-hh beamscreen depicted in Fig.~\ref{FCC_bs}. The final result for $\mathrm{Im} \left( Z_{||} \right) / f$ is the sum of the elements of the third column weighted with the corresponding factors $\alpha$.}
\begin{ruledtabular}
\begin{tabular}{lccr} 
Mode number & $f_{syn}$ [GHz] & $\frac{1}{f_{syn}} \big[ \frac{R}{Q}\big]_{syn}^{||}~[\frac{\Omega}{\mathrm{GHz}}]$ & $\alpha$ \\
\hline
1 & 0.00 & $1.66 \times 10^{-8}$ & 20.3\\
2 & 9.00 & $1.36 \times 10^{-8}$ & 0.85\\
3 & 9.30 & $3.82 \times 10^{-9}$ & 1.35\\
4 & 9.50 & $3.41 \times 10^{-8}$ & 0.68\\
5 & 10.70 & $7.36 \times 10^{-10}$ & 0.61\\
... & & & \\
38 & 27.31 & $1.18 \times 10^{-10}$ & 0.53\\
39 & 27.33 & $1.07 \times 10^{-11}$ & 0.77\\
40 & 27.45 & $3.28 \times 10^{-11}$ & 0.75 \\
\hline
\multicolumn{2}{l}{$\mathrm{Im} \left( Z_{||} \right) / f$ (traveling wave)} & $4.49 \times 10^{-7}~\Omega/ \mathrm{GHz}$ & \\
\end{tabular}
\end{ruledtabular}
\label{FCC_7p5mm_long}
\end{table}

\begin{table}[hbt]
\centering
\caption{Transverse traveling wave data for one period of the FCC-hh beamscreen depicted in Fig.~\ref{FCC_bs}. The final result for $\mathrm{Im} \left( Z_x \right)$ is the sum of the elements of the third column weighted with the corresponding factors $\alpha$.}
\begin{ruledtabular}
\begin{tabular}{lccr} 
Mode number & $f_{syn}$ [GHz] & $\frac{2 \pi f_{syn}}{c} \big[ \frac{R}{Q} \big]_{syn}^x ~[\frac{\Omega}{m}]$ & $\alpha$ \\
\hline
1 & 6.81 & $5.13\times10^{-5}$ & 5.84\\
2 & 9.32 & $5.16\times10^{-5}$ & 0.71\\
3 & 9.43 & $1.20\times10^{-5}$ & 1.15\\
4 & 10.06 & $1.03\times10^{-6}$ & 0.58\\
5 & 10.45 & $7.07\times10^{-6}$ & 2.16\\
... & & & \\
38 & 26.97 & $1.28\times10^{-7}$ & 0.68\\
39 & 27.36 & $2.01\times10^{-7}$ & 0.54\\
40 & 27.87 & $3.99\times10^{-7}$ & 0.53\\
\hline
\multicolumn{2}{l}{$\mathrm{Im} \left( Z_x \right)$ (traveling wave)} & $4.48 \times 10^{-4}~\Omega/m$ & \\
\end{tabular}
\end{ruledtabular}
\label{FCC_7p5mm_trans}
\end{table}

\section{Conclusions}
We have shown that the traveling wave method is well suited for calculation of the longitudinal and transverse impedances of pumping holes.
The method was benchmarked against the CST time domain wakefield solver and the analytical formulas for three different types of geometries.
The method agrees well with both approaches in the cases when they are applicable.
More than that, the method gives converging results even in the cases when one or both of the compared appoaches cannot be applied.
In particular, the analytical approach only works for simple geometries, and only for some sets of parameters (\autoref{circ_pipe_sec}).
The wakefield solver, on the other hand, fails to calculate very low impedances due to the numerical noise (\autoref{FCCbs_sec}).
As a consequence, the traveling wave method is the only option for the FCC-hh beamscreen, where the geometry is complex and the impedance per unit length is very low.
In this case, the method provides a three orders of magnitude better sensitivity than the wakefield solver.

The transverse impedance of the pumping holes in the FCC-hh beamscreen is found to be negligible for single bunch instabilities.
This is a consequence of the hole shielding that reduces the impedance by more than a factor of 30000.

\section{Acknowledgements}
This work was supported by the European Union's Horizon 2020 research and innovation programme under grant No 654305. 
We also thank Walter Wuensch for the useful discussions.

\section*{Appendix: correction factors due to the group velocity}

Below we will show the validity of Eq.~(\ref{Z_equations}), and, specifically, derive the correction factors $\alpha(v_g)$ listed in Eq.~(\ref{alpha_equations}).
To do that, we represent an infinite structure as a limiting case of an $N$-periods long structure, with $N \to \infty$. 
In a finite structure, instead of a continous dispersion curve only discrete points $\phi_p= p \pi / N$ exist (the circles in Fig.~\ref{Dispersions_example}). 
Each point corresponds to a standing wave, and for each standing wave, the resonator impedance model~\cite{Zotter} can be applied:
\begin{equation}
\begin{split}
Z_{||}(\omega) = \frac{R_{||}}{1 + i Q \left( \frac{\omega}{\omega_{res}} - \frac{\omega_{res}}{\omega}\right)}, \\
Z_{\perp}(\omega) = \frac{\omega_{res}}{\omega}\frac{(\omega_{res} / c) R_{\perp}}{1 + i Q \left( \frac{\omega}{\omega_{res}} - \frac{\omega_{res}}{\omega}\right)},
\label{ResonatorFormula}
\end{split}
\end{equation}
where $\omega_{res}$ is the resonant frequency, $Q$ is the quality factor, and $R_{||}$ and $R_{\perp}$ are defined according to Eq.~(\ref{RoverQ}).
In order to make the transverse resonator formula consistent with the adopted definition for $R_{\perp}$ (in $\Omega$), the $R_{\perp}$ (measured in $\Omega / m$ in~\cite{Zotter}) was multiplied by the additional factor $(\omega_{res} / c)$.
By decomposing Eq.~(\ref{ResonatorFormula}) into the powers of $\omega$ and summing over all standing waves (without any correction factors $\alpha_n$), we get
\begin{equation}
\begin{split}
Z_{||}(\omega) = i \omega \sum_{n,p} \frac{1}{\omega_{n,p}} { \left( \frac{R}{Q} \right)_{n,p}^{\Sigma,||}} + O(\omega^2), \\
Z_{\perp}(\omega) = i \sum_{n,p} \frac{\omega_{n,p}}{c} { \left( \frac{R}{Q} \right)_{n,p}^{\Sigma,\perp} } + O(\omega).
\label{ZvsRoverQstanding}
\end{split}
\end{equation}
Here the symbol $\Sigma$ means that the shunt impedances are not per-period, but the total impedances of the structure. 
We will show that the sum over the indexes $n \times p$ of the standing waves can be reduced to the sum over indexes $n$ of the synchronous traveling waves with the correction factors $\alpha_n(v_g)$, as in Eq.~\ref{Z_equations}.

Let us consider only one mode $n$ and omit the index $n$ below.
To calculate the shunt impedances $\left( R/Q \right)_p^{\Sigma}$, we first write down the total voltage kick $V_p^{\Sigma}$ (either longitudinal or transverse). 
The electromagnetic force $F$ of a standing wave can be expressed as a sum of the forward and the backward waves
\begin{equation}
\begin{split}
V_p^{\Sigma} = \frac{1}{e} \int_0^{NL} \left( F^++F^- \right)|_{s, t=s/c} ds=\\
= \frac{1}{e} \int_0^{NL} F_0(s) e^{i \omega_p s/c} \left( e^{-i \phi_p s/L} + e^{i \phi_p s/L} \right) ds
\end{split}
\end{equation}
where $e$ is the elementary charge and $F_0(s)=F_0(s+L)$ is the periodic envelope function for both the forward and the backward waves, and the index $p$ can be any integer in the range $0 \leq p \leq N-1$. 
Using the periodicity, we write
\begin{equation}
\begin{split}
V_p^{\Sigma} = \bigg[ \int_0^{L} \frac{F_0(s)}{e} e^{i \omega_p s/c-i \phi_p s / L} ds \bigg] \times \\
\times \left(1+e^{i (L\omega_p/c-\phi_p)}+e^{2i (L\omega_n/c-\phi_p)}+...\right)+\\
+ \bigg[ \int_0^{L} \frac{F_0(s)}{e} e^{i \omega_p s/c+i \phi_p s / L} ds \bigg] \times \\
\times \left(1+e^{i (L\omega_p/c+\phi_p)}+e^{2i (L\omega_p/c+\phi_p)}+...\right),
\end{split}
\label{long_eq_for_V}
\end{equation}
where the first summand corresponds to the $+p$ branch and the second summand corresponds to the $-p$ branch in  Fig.~\ref{Dispersions_example}.
Futhermore, we can assume that for each $p$, only one of the branches matters.
Indeed, if the considered wave $(\phi_p, \omega_p)$ lies close to the synchronous line $\omega = \phi c / L$ on the dispersion diagram, its counter-part $(-\phi_p, \omega_p)$ will be far from synchronous and will not amount to a significant impedance.
Therefore, we can simplify Eq.~(\ref{long_eq_for_V}) by leaving only the first summand, and exdending the range of $p$ to $-(N-1) \leq p \leq N-1$.

\begin{figure}[!htb]
\centering
\includegraphics[width=90mm, angle=0]{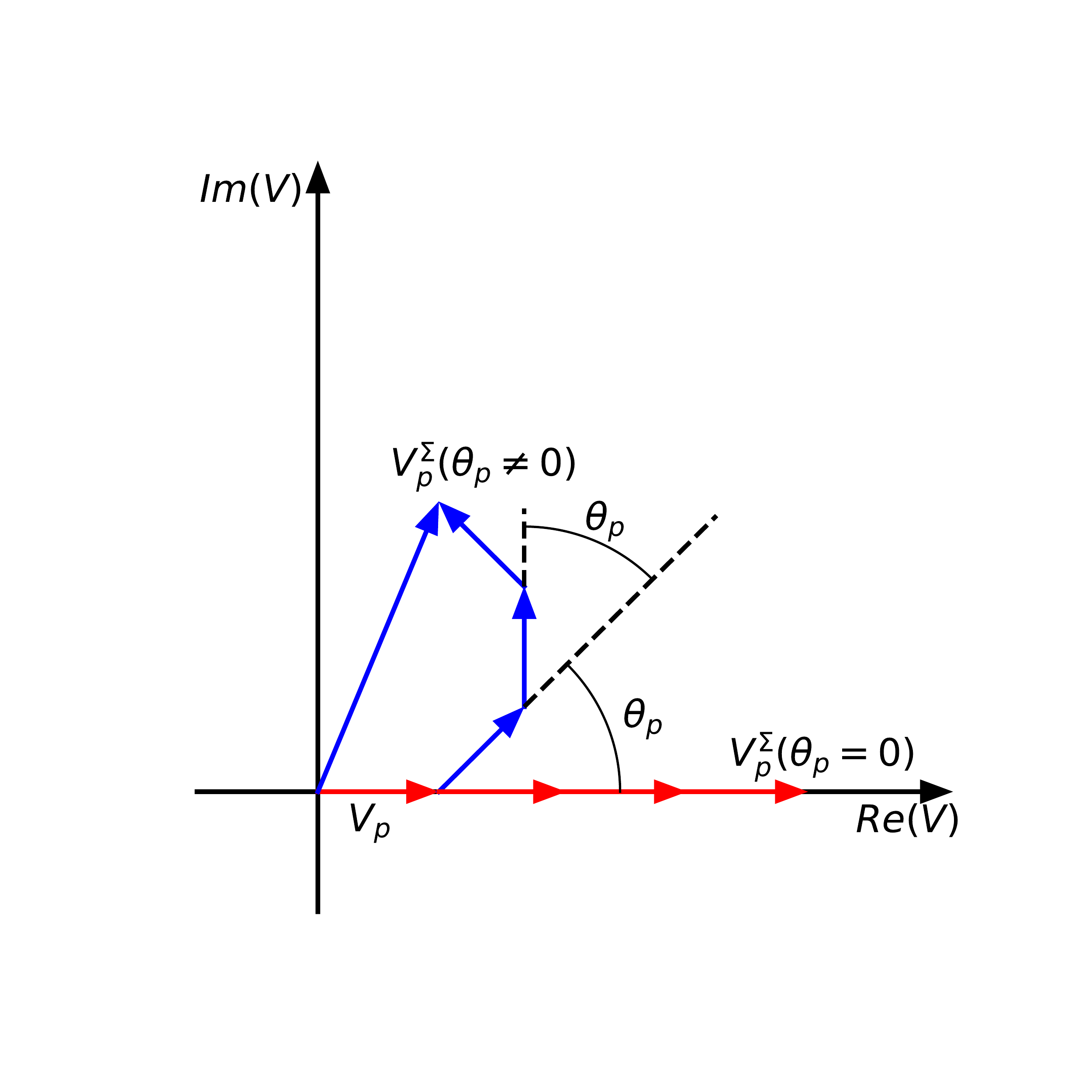}
\caption{Geometrical representation of the voltage summation $N=4$ periods of the structure. The perfectly synchronous case is shown in red, and an asynchronous case is shown in blue.}
\label{V_sum}
\end{figure}

The integral in the first square brackets of Eq.~(\ref{long_eq_for_V}) is equal to $V_p$ - the voltage of the $p$-th traveling wave in a structure consisting of only one period. Therefore
\begin{align}
V_p^{\Sigma} \approx V_p (1+e^{i \theta_p}+e^{2i \theta_p}+...+e^{2i (N-1)\theta_p}),
\label{V_sum_eq}
\end{align}
where $\theta_p=L\omega_p/c-\phi_p$ is the phase slippage per period.
A geometrical representation of Eq.~\ref{V_sum_eq} is depicted in Fig.~\ref{V_sum}.
The farther away the point $p$ is from the synchronous point (a cross in Fig.~\ref{Dispersions_example}), the lower is its total voltage.
Mathematically it can be expressed as $|V_p^{\Sigma}| = N |V_p| g_N(\theta_p)$, where the function $g_N(\theta)$ describes the attenuation of the voltage sum due to asynchronicity.
From Eq.~(\ref{V_sum_eq}) it can be shown that
\begin{equation}
g_N(\theta) = \frac{\mathrm{sin}(N \theta/2)}{N \mathrm{sin}(\theta/2)}.
\end{equation}
This compact result for $|V_p^{\Sigma}|$ is only possible thanks to the assumption that only one of the forward and the backward waves matters. 
If both waves were considered, $|V_p^{\Sigma}|$ would have to depend on the two voltages as $|V^+ + V^-|$, where $V^+$ and $V^-$ correspond to the terms in Eq.~(\ref{long_eq_for_V}).

In one period of the structure, the energy stored in a standing wave is twice that stored in the corresponding traveling wave. 
Over $N$ periods, the energy of the standing wave becomes $U_p^{\Sigma} =2NU_p$.
Given $|V_p^{\Sigma}|$ and $U_p^{\Sigma}$, the standing wave shunt impedance can be related to the per-period impedance of $p$-th traveling wave:
\begin{equation}
\left( \frac{R}{Q} \right)_p^{\Sigma} = \frac{|V_p^{\Sigma}|^2}{2 \omega_p U_p^{\Sigma}} = \frac{N}{2} \left( \frac{R}{Q} \right)_p g_N^2(\theta_p).
\end{equation}
We then plug this expression in Eq.~\ref{ZvsRoverQstanding} and take the limit $N \to \infty$ to obtain the contribution of the $n$-th resonance to the total impedances
\begin{equation}
\begin{split}
\frac{Z_{||}(\omega)}{N} = i \omega \lim_{N \to \infty} \sum_p \frac{1}{\omega_{p}} \left( \frac{R}{Q} \right)_p^{||} \frac{g_N^2(\theta_p)}{2} + O(\omega^2), \\
\frac{Z_{\perp}(\omega)}{N} = i \lim_{N \to \infty} \sum_p \frac{\omega_p}{c} \left( \frac{R}{Q} \right)_p^{\perp} \frac{g_N^2(\theta_p)}{2} + O(\omega). \\
\label{intermediate_Z}
\end{split}
\end{equation}
As $N$ increases, the function $g_N(\theta)$ becomes a sharper and sharper peak around $\theta = 0$.
Using this fact, the quantities $(R/Q)_p^{||} \omega_p^{-1}$ and $(R/Q)_p^{\perp} \omega_p$ can be taken out of the summation and replaced with $(R/Q)_{syn}^{||} \omega_{syn}^{-1}$ and $(R/Q)_{syn}^{\perp} \omega_{syn}$, respectively.
In the important case when $\omega_{syn} = 0$, a limit has to be taken while approaching the synchronous point: $\lim_{\omega' \to 0}(R/Q)^{||}(\omega') / \omega'$.

After bringing back the summation over the bands $n$, Eq.~(\ref{intermediate_Z}) transforms into the sought-after Eq.~(\ref{Z_equations}), with
\begin{equation}
\alpha(v_g) \equiv \lim_{N \to \infty} \sum_{p=-(N-1)}^{N-1} \frac{g_N^2(\theta_p)}{2}.
\label{alpha_and_gN}
\end{equation}
The only remaining part is to prove that the coefficient $\alpha(v_g)$ can be written as a function of the group velocity in the form of Eq.~(\ref{alpha_equations}).
To do that, the cases $\omega_{syn} \neq 0$ and $\omega_{syn} = 0$ are treated separately.

Let us first treat the case $\omega_{syn} \neq 0$.
In the vicinity of the synchronous point, the dispersion curve can be approximated by a straight line $\omega_p = \omega_{syn} + (\phi_p-\phi_{syn}) \partial \omega / \partial \phi = \phi_{syn} c / L +(\phi_p-\phi_{syn}) v_g / L$.
The phase slippage per period becomes
\begin{equation}
\theta_p = \left( 1-\frac{v_g}{c} \right) (\phi_{syn} - \phi_p).
\label{theta_p_def}
\end{equation}
The coefficient $\alpha(v_g)$ can be expanded as
\begin{equation}
\alpha = \lim_{N \to \infty} \frac{1}{2} \sum_{p=-(N-1)}^{N-1} \bigg[ \frac{\mathrm{sin}(\xi (p^*-p))}{N \mathrm{sin}(\frac{\xi}{N}(p^*-p))} \bigg]^2
\label{alphaxi}
\end{equation}
where we have defined $\xi=\frac{\pi}{2}(1 - v_g / c)$ and $p^*=\phi_{syn} N / \pi$.
As $N$ increases, the terms in the denominator approach $\xi (p^*-p)$, and $\alpha$ approches
\begin{equation}
\alpha = \frac{1}{2} \sum_{p=-\infty}^{\infty} \mathrm{sinc}^2(\xi (p^*-p))
\end{equation}
We can then use a remarkable property of the sinc function $\sum_{k=-\infty}^{\infty} \mathrm{sinc}^2(a k - b)=\sum_{k = -\infty}^{\infty} \mathrm{sinc}^2(a k) = \pi / a$ and rewrite $\alpha$ in a more compact way
\begin{equation}
\alpha = \frac{\pi}{2 \xi} = \frac{1}{1-v_g/c}
\end{equation}

Now, let us consider the remaining case $\omega_{syn} = 0$. 
The sum in Eq.~(\ref{alpha_and_gN}) has to be split in two parts because for the negative $p$ the group velocity has the opposite sign $v_g(-\phi_p) = -v_g(\phi_p)$, as shown in Figure~\ref{Dispersions_example}.
By following the same steps as for the case $\omega_{syn} \neq 0$, we arrive to
\begin{equation}
\alpha = \frac{1}{2(1-v_g/c)} + \frac{1}{2(1+v_g/c)} = \frac{1}{1-v_g^2/c^2},
\end{equation}
which proves Eq.~(\ref{alpha_equations}).

\end{document}